# Blockchain scalability for smart contract systems using eUTXO model

Cayo Fletcher-Smith [*], Frazer Chard [†]

[*]Bournemouth University

[*]s5117135, s5115137@bournemouth.ac.uk

*Abstract*—This research critically analyses blockchain scaling solutions based on their ability to realistically balance properties of the blockchain trilemma. We've concluded this research by outlining a gap in the current body of literature, and implementation of scalability solutions. An extended-UTXO transaction model is proposed to overcome challenges associated with implementing both layer-one and layer-two scaling solutions in a blockchain system. The examination of industry approaches is used to justify this direction, and puts forth a basis for future work.

## I. INTRODUCTION

Blockchain technology was first introduced by Satoshi Nakamoto in the Bitcoin whitepaper, in which they proposed a distributed ledger of transactions stored across a peer-to-peer (p2p) network of nodes. Transactions facilitate the transmission of digital assets over the blockchain network and store a record of that transaction in blocks using cryptographic trust to verify data validity, circumventing the need for a trusted central authority. Each block can be visualised as a single page, while the entire blockchain would be a full book [1]. Blockchain use has been growing exponentially since Bitcoin's rise in 2013, and Ethereum's in 2017. Smart contract projects like Solana and Cardano are pushing the boundaries encompassed in Web2, shifting design away from centralised infrastructure, by providing disruptive solutions to business sectors globally. The magnitude of this shift in thinking has accelerated a domain of decentralized financial services, applications and industry solutions due to the inherent benefits blockchain technology has across various use cases. Despite significant advancements since the days of primitive input-output ledger systems like Bitcoin, the industry still struggles to correctly balance the coexistence of features fundamental to a comprehensive blockchain solution.

### A. Blockchain Trillema

Three fundamental properties of blockchain technology: decentralisation, security and scalability are regularly observed in research to have a challenging relationship. Generally to achieve two properties in-full projects sacrifice the third in some form. Zhou et al exemplifies the trilemma by explaining that, an introduction of centralised coordination can increase transaction speed by reducing computation costs, resulting in the sacrifice of complete decentralisation [2].

### B. Research Objectives

In this research, we will be exploring recent technological developments in scaling at different layers of the blockchain stack and perform a comparative analysis to review how effectively the blockchain trilemma can be balanced with these solutions.

Observations will be made throughout about the feasibility of certain scaling solutions occurring harmoniously with the layer-one transaction models implemented across an array of projects. The necessity and support for smart contracts will be demonstrated where appropriate, both in transactional model support and automated scalability requirements.

## II. BACKGROUND

### A. Ledger State

At a high level, blockchains can be seen as state machines, which are used to permanently and immutably record the change in the state of the ledger. A state machine is a mathematical model which represents a machine with exactly one state at any time, capable of transitioning to a new state based on external interactions and documenting previous states [3]. The two main models used to represent how the state is recorded are Unspent Transaction Outputs (UTXO) and Account-based The UTXO model performs local and deterministic state management through the set of all transaction outputs stored within a ledger and is most notably used by Bitcoin. Following a newly created transaction, UTXOs are generated from the output to represent the value transferred and are then assigned to the receiver's public address. A user can use any UTXOs that are assigned to their private key for further transactions, but once the transaction is verified and stored in the ledger, the UTXO will be considered spent and is no longer usable. This model does not incorporate accounts at the protocol level but uses wallets to maintain balance by recording the sum of all UTXOs for an account [4]. The account-based model tracks the global shared state of asset balances for all accounts on the network, with the Ethereum blockchain following this state model. The two account types seen in this model are the private-key user accounts and global smart contract accounts, both of which contain executable code, account balance and internal memory. When a transaction is sent and verified on the ledger, the sender's balance is decreased and the receiving account balance is increased by the same value. After a transaction has occurred, the global state is updated to reflect any changes to account balance [5]. The UTXO model maintains simple and deterministic state in distributed and concurrency computing environments, but is limited in functionality to execute expressive scripts. In contrast, the account ledger is focused on completely expressive smart contracts on a global shared state. IOHK proposed the Extended UTXO model, which builds upon Bitcoin's UTXO model to maintain the inherent semantic simplicity while supporting a more expressive form of smart contracts than UTXO allows [6].

### B. Public Key Cryptography

Public key cryptography (PKC) is built on the cryptographic primitive of Trapdoor Functions, which is a mathematical function that can be solved simply in one direction but is almost impossible to reverse and is integral to every single transaction that occurs on the blockchain [7]. Public key cryptography is used to create an algorithmically linked public-private key pair, comparable to a username and password. The private key is used to authorise and sign transactions being sent from the user, creating digital signatures that enforce sender provenance [8]. The public key is used to encrypt a transaction before being broadcast to the entire network to enhance privacy and confidentiality assuring that only the intended private key owner can decrypt and view the data. The keys and public addresses are managed by a wallet, either an application or specialised hardware, that keeps track of private key balances [8].

### C. Blockchain Transactions

In both the UTXO and account-model transaction systems, sending digital assets to another account on the network follows very similar principles of transaction [9]. Each transaction contains a message with confirmation of both Alice the sender and of Bob the receiver, as well as the value of the digital assets being transacted. Alice will then encrypt and sign the transaction with the relevant keys, and then broadcast the transaction along with a fee to the p2p blockchain network [10]. Validator nodes will select transactions to be validated and generally prefer to maximise revenue from transaction fees. After the block is broadcast, validator nodes will validate that the transaction is viable and accurate then package it into a block with other unconfirmed transactions. This is then broadcast to all other network participants to ensure that the transaction inputs are signed and have not been already spent. When other nodes agree on the validity of a block they will receive a small incentive and the block will be passed along and queued for inclusion in the ledger, eventually coming to a global consensus when enough validators have confirmed the block to be valid [11]. Once the block has been added to the blockchain and confirmed, Bob will then receive the digital asset sent by Alice.

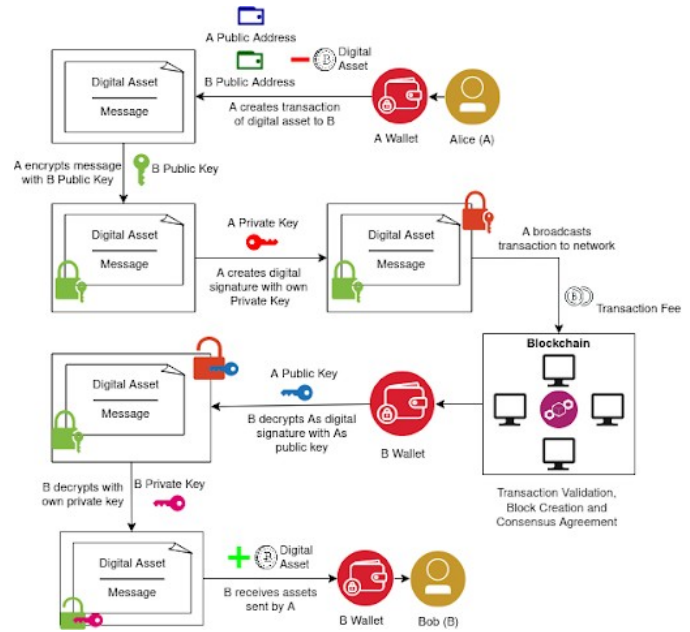

Fig. 1: Blockchain Transactions

### D. Smart Contract

Smart contracts are pieces of coded logic stored within a blockchain network in the form of high-level contract-oriented languages to be executed by nodes either manually or when certain conditions are met. Smart contracts contain state storage, specialised logic, a balance and a public/private key address and are capable of receiving assets. Contracts have been integrated into many blockchain networks and serve as part of the core technology enabling transactions to occur in decentralised applications [8]. The implementation of contracts allows more complex transactions of valued assets to occur without censorship, downtime or any centralised third parties to oversee transactions [9] [12].

### E. Consensus Mechanism

Consensus mechanisms ensure that nodes within a distributed network stay synchronised and follow a consistent set of rules on how to validate new blocks and achieve Byzantine Fault Tolerance (BFT). BFT is a property of distributed systems that enable the participating nodes to agree on a single strategy to achieve global consensus regardless of failed components or untrustworthy nodes. These mechanisms enable blockchain systems to act in a trustless and decentralised manner, especially when interacting with faulty or malicious nodes [13]. Consensus mechanisms enable large groups of validator nodes to vote upon which block of transactions should be validated in a process called minting [13]. After a block has been minted by a validator and confirmed to coincide with the copy of the ledger currently maintained by them, it is appended to their copy of the ledger and broadcast out, coming to a global consensus on the state of transactions. The weight of a validators vote is proportional to the quantity of resources contributed to the network, with proof-of-work requiring computational power and the proof-of-stake consensus relying on tokens pledged to the network.

The requirement of resources to achieve consensus is deemed essential, as a "one-node, one-vote" approach could result in many nodes being set up to spoof the network and receive an imbalance of voting power [14].

## III. RELATED WORKS

This section examines the various approaches to blockchain scaling, at each level of the blockchain stack. Layer-one consists of on-chain solutions, which is where the transactional model, consensus mechanism and blockchain architecture sits. Layer-two consists of any off-chain solution used to support the operation of an underlying layer-one solution, this includes sidechains and off-chain processing.

### F. Blockchain Structure

Once consensus has been achieved and there have been enough confirmations validating the transactions within the block, the block will be appended to the ledger and broadcast to all nodes to ensure a consistent state. Each block contains a timestamp for when it was appended to the chain, a random nonce (number used once), a merkleized transaction root hash, transaction data, the hash of the previous block and the hash of the current block, which is generated by hashing everything else contained within the block [15]. The previous block's hash is tethered using a hash pointer, linking each block together in an immutable chain, seen in figure 2. If anything in a previous block is altered it will invalidate the entire chain from that point forward [15].

### A. Layer 1 - On-chain

Transaction throughput doesn't scale alongside network growth due to processing occurring sequentially. To solve this, the layer 1 network can be divided into shards acting as singular entities processing transaction validation in parallel. This allows throughput to scale out as the network grows (see Fig. 4). The following explores how select challenges are approached in different projects and literature.

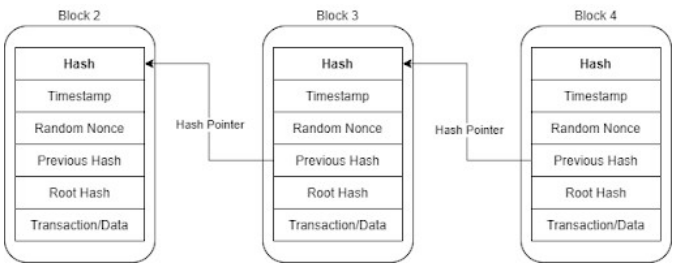

Fig. 2: Blockchain Structure

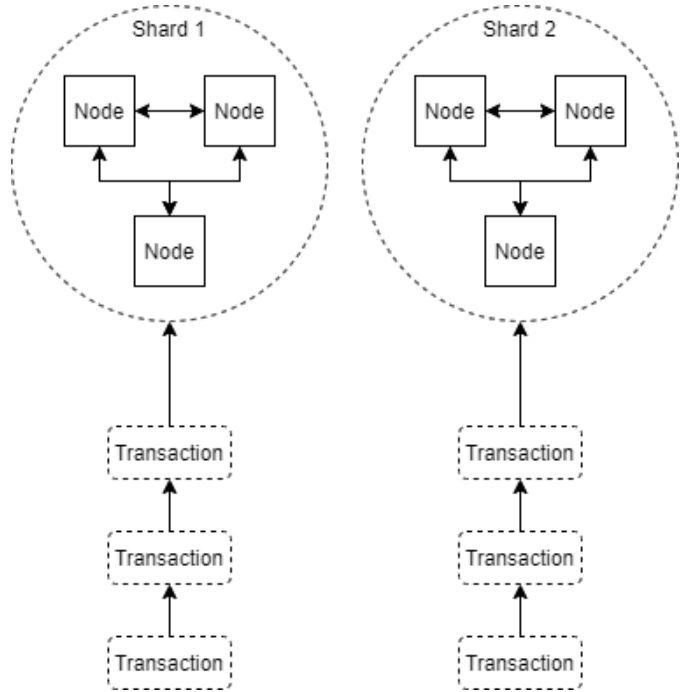

Fig. 4: Sharding

### G. Merkle Root Tree

One of the key cryptographic functions utilised in blockchain technology is the Merkle root hash, which is a hash that represents all transactions within a block. The hash values of each transaction are appended together and then further hashed together until a single root hash is formed, as seen in figure 3 [16] [17]. If a singular bit of transaction data is changed within the Merkle tree, the root hash would no longer be valid and the chain would break. This method ensures an immutable and integrity-focused system where any alteration in block data will lead to inconsistency when validating the ledger [11].

*1) Network Sharding:* The first challenge in sharding is securely segmenting the network into smaller groups. Each shard must maintain Byzantine fault tolerance, which is difficult as shards are exponentially smaller than an entire network. To ensure probabilistic Byzantine fault-tolerance, shard assignment must be unbiased and randomised. Elastico tackles this in their identity establishment and shard formation phase, at the start of each epoch [18]. This is where nodes each generate an identity based on a correct proof-of-work (PoW) solution. The PoW solution is a publicly verifiable hash and has the following values inserted into it: "epoch randomness", "IP address", "public key" and "nonce". Nodes are assigned uniformly at random to shards based on specific bits in the output hash of the PoW function. For example, all nodes with hashes ending in "0010" are assigned to "shard 1" and all hashes ending in "0111" to "shard 2". Elastico assures epoch randomness is generated by a designated "final shard" at the end of each epoch, for use in the identity establishment of the next. Final consensus is achieved by the "final shard", which computes the merkelized value of blocks from all other shards. The result of this is broadcast across the entire network at the end of each epoch [18]. Instead of Elastico's PoW based identity establishment, OmniLedger uses a verifiable random function (VRF) to generate a ticket used in the unbiased and unpredictable election of a temporary shard coordinator. The VRF is a keyed cryptographic hash generated using a node's private key, that can be verified and accessed using their public key [19]. In the event the elected coordinator is deemed to be malicious, OmniLedger uses the RandHound protocol to generate an unbiasable random string (URS) calculated from the previously created VRF of each node. The URS is computed by the temporary leader, but the output is completely dependent on the VRFs produced by each individual node [20].

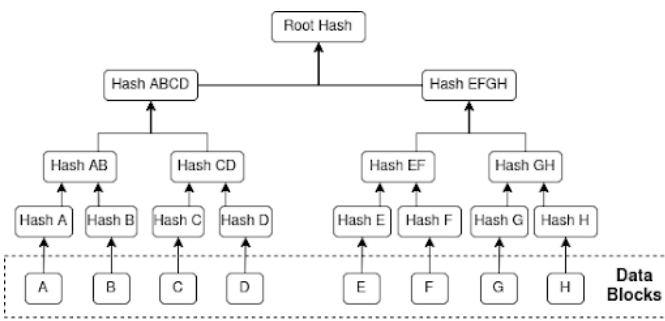

Fig. 3: Merkle Root Hash

*2) Transaction Sharding:* After shard assignment, the next challenge is determining how transactions will be allocated to individual shards to ensure that double-spending isn't possible by transacting the same funds in multiple shards [21]. Transactions can instead be handled based on certain attributes such as their sending address [22]. An address's subsequent transactions are therefore handled by the shard which stores the transaction originators current UTXO state. Zilliqa's sharding protocol uses a single directory service shard (DSS), to assign transactions to processing shards based on attributes, which operates similar to Elastico's "final shard". The DSS is also used in block synchronisation at the end of each epoch and can be conceptualised as a coordinator similar to beacon/parent chains used in layer 2 approaches [22].

*3) Cross-shard Transactions:* Finally, cross-shard transactions are critical to ensure the validity of transactions complying with different changing UTXO states across different shards [21]. OmniLedger proposes an atomic commit protocol to manage the finalisation, or roll back transactions based on the consensus of multiple shards [20]. Atomic commit transactions occur in three phases: (i) initialization, (ii) lock and (iii) unlock [21]. In this protocol, the client (transaction originator) is leveraged as an atomic coordinator. In initialization, the transaction is sent to multiple shards where the UTXO state is locked. Shards individually achieve consensus, responding with a cryptographic proof-of-acceptance (PoA) or proof-of-rejection (PoR) hash to the client. To commit the transaction, the client sends all PoAs to a predefined shard, where the transaction is unlocked and committed. Similarly, if any PoR is received, the client sends this to shards maintaining the locked UTXO, where it's unlocked and rolled back [20]. This process ensures the transaction is either committed to a block on one shard or unlocked and rolled back on all shards. If the client doesn't respond the state will be unlocked and rolled back allowing the UTXO to be spent in another transaction [21].

### B. Layer 2 - Off-chain

Off-chain solutions are used to simulate horizontal scaling found with layer 1 sharding techniques. Alternative to small network shards, sidechains are used to handle transactions, achieving similar parallelization to individual UTXO shards. Sidechains rely on different validators and mechanisms to process transactions easing on-chain congestion. Such chains use two-way pegs to provide interoperability between the mainchain being scaled and itself, allowing a multidirectional asset transferral The mainchain adopts a parent role, dictating underlying security, approving off-chain processing, and arbitrating off-chain disputes. This is comparable to designated coordinators in some layer-one sharding protocols [23]. Generally, the UTXO model is used to easily represent whether a state has been spent or not, ensuring accurate state transitions of the main-chain while minimising mass-exit costs. In Fig. 5, the two-way peg is seen sitting between the parent and child-chain. Tokens from the parent are locked, with the equivalent value being created/unlocked on the child for the transaction to be processed [24].

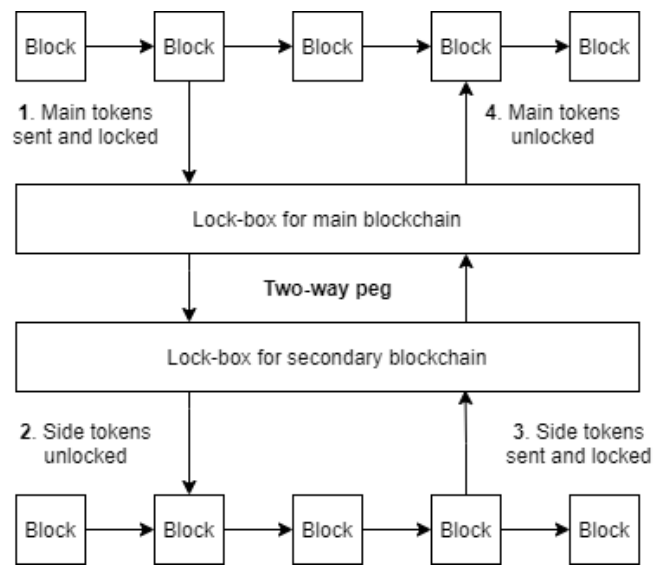

Fig. 5: Sidechain two-way peg

Sidechain networks can be created, with a tree-like recursive parent/child relationship, filtering transaction processing away from bottlenecks. This approach to parallelization is known as a nested-chain. Transactions are delegated to children, which in turn may have generations of chains under them. This circumstance requires a chain to act in both a child and parent role. This architecture is implemented in the Plasma network, to increase Ethereum's scalability and is demonstrated in Fig. 6 [25].

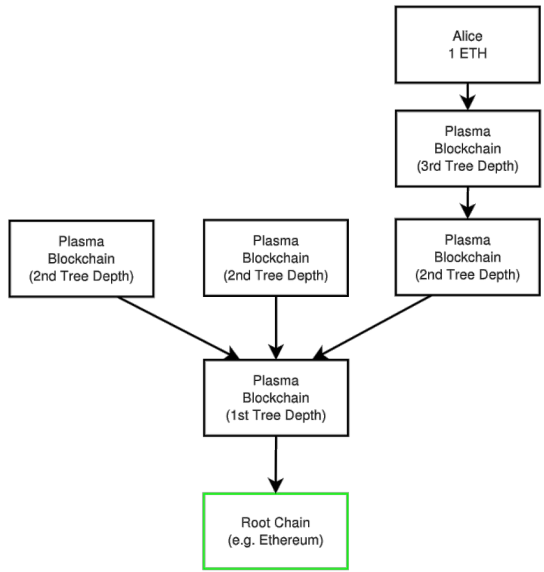

Fig. 6: Plasma Network - Nested Chains

Upon the processing of transactions, child-chains transform all transactions in the block into a single cryptographic value and then periodically record the results back to the parent. Such processing can be specialised to the sidechain, for example, Bitcoin Lightning Network handles microtransactions for Bitcoin, saving bandwidth for large security-critical transactions on the main chain [26]. Plasma network scales the execution of smart contracts, made possible by Ethereum's account-based transaction model. Smart contracts automate the offload of transactions, to specific Plasma chains based on predefined transaction attributes. Plasma processes these and submits the merkleized fraud-proof back to Ethereum [25]. This enables users to report any dishonest nodes to their parent-chain, roll back the block to protect digital assets and penalise the malicious actor, which helps protect against the risk of Sybil attacks and other forms of corruption that may occur [27].

## C. Layer 2 - Hybrid Scaling

Rollups are a recent development in scaling blockchain technologies which are considered to be a hybridization of both layer 1 and layer 2 solutions [27]. Rollups work in a similar manner to nested-chains but are trusted to scale more by solving specific data availability issues that are seen with complete layer 2 solutions, rather than just moving all data, storage state and computation onto a side-chain. Transactions that are about to be processed are "rolled up" into a single batched transaction, and then compressed to decrease the size and cost of each batch, enabling more efficient transactions to occur. Rollups only migrate the transaction execution and state storage off-chain into specialised "rollup" contracts while retaining just enough data about each transaction on the main-chain [26].

Once the execution is complete, the results of the change in state will be relayed back to layer 1. Storing transaction data on-chain is essential to maintain transactional security as it enables the integration of a byzantine-fault tolerant consensus mechanism to manage the trustless state of consensus on the layer 1 chain [26]. By securing rollups with blockchain security mechanisms and storing transaction data on-chain, the hybrid system allows anyone user to process transactions locally to detect fraud, initiate withdrawals from the rollup and also start producing batches of transactions.

Each rollup stored on-chain maintains a Merkle root hash of the batched transactions for the current state of the contract, known as the state-root, which includes account balances and rollup code logic. This is vital when verifying the state of a batch to see if the previous state-root hash of the batch matches with the current root-state. If they do match, the proposed state-root will transition into a new rollup state-root [27].

Although the transaction data stored on-chain is secured through consensus mechanisms, the transactions being executed and the change of state within the rollup contract will not be validated by the main-chains consensus. Rollups require alternative mechanisms to prove that the state-root of the contract following batch transaction confirmation is accurate. This is needed as without batch verification, a malicious actor can submit a batch with a fraudulent Merkle-root and incur no consequences, then being capable of transferring all digital assets within the rollups to their own account. There are two completely different approaches to solve the issue of fraudulent states, which are optimistic rollups and zero-knowledge rollups [28].

*1) Optimistic Rollups:* Optimistic rollups follow an interactive approach to verify whether a submitted batch has a fraudulent post-state root. Each rollup contract tracks the historical state-roots and the hash of each transaction batch. Transaction batches are confirmed optimistically, but if users discover that one of the batches produces an incorrect post-state root, they can generate and publish a fraud-proof containing the transaction and part of the Merkle tree to the chain proving that the batch was processed incorrectly [29]. Rollup contracts verify whether the fraud-proof is legitimate, and if it is deemed to be, that batch and all consequent batches are rolled-back. Rollups benefit from the main-chain security but do offer a lower throughput than Plasma and ZK Rollups. The trade-off for the lower throughput is the ability to scale smart contracts with an optimistic model [26]. Optimistic rollup projects include Optimism and Arbitrum [30] [31].

*2) Zero-Knowledge Proofs:* ZK-rollups solutions are based on the concept of zero-knowledge proofs and validity proofs, which are both cryptographic proofs that get included with every batch. The ZK rollup framework makes use of zero-knowledge proofs, which lets observers immediately prove the validity of an assertion by ensuring that the post-state recorded to the main chain is the accurate result hash from executing the batch. This can be verified very quickly on-chain, regardless of how big the computation is, however, the computation required to create a ZK proof is expensive and only suitable for transaction management and not complex smart contract execution [26]. There are projects which are attempting to explore ZK-Rollups, including ZKSync which uses the SNARK implementation and StarkEX implementing STARK proofs [32] [33].

Once a user wishes to withdraw their funds from the child-chain back to the main-chain, the user balance is calculated using spent and unspent transaction outputs that occurred on the side-chain associated with the user's private key. This value of the native tokens on the side-chain is then put into a lockbox and unlocked on the parent-chain. There is typically a validation period for confirmation of the spent and unspent transactions to ensure accuracy [34].

## IV. Comparative Analysis

This section examines the impact on each property of the blockchain trilemma when implementing layer-one and layer-two scaling solutions across UTXO and account-based transaction models. Comparable metrics are up-scale factors, transaction throughput and dispute resolution. Qualitative comparison will be used to express the strengths and limitations of decentralisation and security.

### A. Layer-One Sharding

The works of Luu et al. in Elastico presented a decentralised scalable solution to sharding and parallelisation, proposing feasible approaches to shard assignment without central coordination [18]. Although scaling increases with network growth linearly, individual shard security is sacrificed. This culminates in an 8% average shard failure probability, having negative security implications and undermining scaling through failure bottlenecks [21]. Resilience to byzantine threats decreases consistently, as shards are added, meaning infinite scaling eventually completely sacrifices system integrity. This can be partly attributed to small shard sizes [20]. Kokris-Kogias et al address bottleneck and security issues native to Elastico scaling through their work on OmniLedger. By increasing shard size byzantine fault tolerance is more effectively managed in a decentralised setting. The implementation of an atomic commit protocol directly combats double spending, an issue imperfectly solved in industry approaches previous to OmniLedger. Cross-shard atomicity ensures UTXO funds aren't indefinitely locked as pending in-cases of malicious coordination. This was achieved through rejection and acceptance proof, with resilient commit and rollback parameters [20]. Probabilistically, OmniLedger could survive for 230 years before complete shard failure undermines the blockchain. In contract Elastico can survive 1 hour, under the same parameters [35].

Zilliqa utilises directory service shards to increase security and transaction efficiency through transaction coordination This mitigates double-spending and is a scalable solution to transaction sharding. Transaction speeds don't scale as highly compared to RapidChain or OmniLedger, however transaction assignment is more secure[36].

| | Transaction Speed | Linear Scaling | Security | Decentralisation | Shard Resilience |
|---|---|---|---|---|---|
| Elastico | 40 Tx/S | Yes | High shard failure with potential for double spend | No trusted third parties. Weak shards represented distributed single points of failure. | < 33% |
| OmniLedger | 3500 Tx/S | Yes | Robust shard structure and secure transacting. | No trusted third parties or single points of failure. | 33% |
| Zilliqa | 2500 Tx/S | Yes | Transaction assignment assures security objectives. | No trusted third parties or single points of failure. | 33% |

Fig. 7: On-chain sharding comparison

### B. Layer-Two Scalability

Plasma follows a nested, parent/child chain architecture built on Ethereum [9]. Although Ethereum supports a high level of decentralisation, Plasma can centralise aspects of processing activities due to centric coordination on some sidechains [25]. Security challenges are found in consensus, weakening the overall byzantine fault tolerance of the network. For these reasons Plasma severely sacrifices security and decentralisation in the pursuit of scalability.

Compared to Plasma ZKsync achieves far greater decentralisation and security at scale. Despite significantly lower transaction speeds and latency, ZKsync manages all three attributes of the blockchain trilemma while achieving impactful scalability. The zero-knowledge nature of this solution is privacy preserving by default, increasing overall system security. ZKsync is fully dependent on smart contract implementation to aggregate transaction processing, making it infeasible for UTXO based projects [32]. Nested-chains like Plasma can be deployed over a wider array of layer-one implementations, making them more feasible for less specialised use-cases.

Contrasting ZKsync and zero-knowledge rollups in general, optimistic roll ups like Arbitrum focus on hyper-scalability, with disregard for other decentralisation. Supporting the highest scaling of any solution at both layer-one and two, Arbitrum is a prime example of the trilemma. The use of trusted centralised sequencers, undermines the decentralised aims of all permissionless blockchain projects. Security is increased through fraud proofs, although slow generation increases transaction latency. Although transactions are processed quickly, latency causes large time windows until a transaction is immutably confirmed.

| | Plasma | Arbitrum | ZkSync |
|---|---|---|---|
| Architecture | Nested parent/child-chains | Optimistic Rollup | ZKRollup |
| Transaction Speed | 1000+ Tx/S | 40,000 Tx/S | 300+Tx/S |
| Scalability | Child-chains perform work and results are relayed to parent-chain. | Enables higher transaction capacity. | Batches of transactions are performed in a rollup contract. |
| Decentralisation | Child-chains are generally controlled by central groups. | Uses a trusted centralized sequencers who control the ordering of transactions. | Non-custodial validators, have no power over user assets and are not required to withdraw any. |
| Security | Poor consensus mechanisms. | Main consensus, Generating a fraud proof takes typically a week. | Main consensus, cutting-edge cryptographic primitives. |
| Dispute Resolution | Proof of Transaction Spendability & Proof of State Transition | Multi round approach, executing small L2 transaction chunks to find the frauded one. | Transaction is verified by smart contracts through verifying a blocks proof-of validity. |

Fig. 8: Off-chain scaling comparison

## C. State Transaction Models

The core UTXO account model used by Bitcoin was designed to handle transactions from one user to anothers private key address. The way the unspent transactions are generated and spent means that these transactions have a greater chance of being scaled by processing them in parallel. Although the UTXO model has minimal expressiveness in terms of on-chain logic, the UTXO model enables scalability and confidentiality, with much simpler transaction verification [11].

Vitalik introduced the Account-model into the Etheruem blockchain, which runs on a global virtual machine which maintains a record of all transactions, users and even smart contracts. Accounts are ran by private keys or smart contracts and store user assets as account balance. Although storing account state within the global state makes things easier, it also limits scalability options at layer one and requires the introduction of layer 2 solutions such as Plasma [9].

More recently, Cardano's IOHK developed the eUTXO model, which maintains the core aspect of the UTXO model with regards to how each transaction consumes unspent outputs in order to produce new unspent transactions that can be further spent. As Cardano intended to implement smart contract functionality, using the basic UTXO model that was designed to do little more than handle payments would be unsuitable. This is highlighted through the lack of expressive coded logic available on-chain to UTXO systems. The use of an Account model addresses the issue of limited expressiveness by developing specialised smart contract accounts, becoming far to complex [37].

To create an eUTXO, two core additional pieces of functionality are needed to build upon the base UTXO model. The model needed to ensure that smart contract state was maintained and also enforce that the same smart contract logic is used throughout the whole sequence of transactions. This eUTXO model enables native-assets and smart contracts to be used on the UTXO model for the first time [6].

| | State Type | User account | Scalability | Security | Decentralisation | Smart Contract |
|---|---|---|---|---|---|---|
| UTXO | Local and deterministic state. | Private-Key Wallet address | Outputs can be easily processed in parallel. | Public key transactions | Implement strong Nakamoto-Style consensus | Not expressive. E.g. Bitcoin Script. |
| Account | Global shared state. | Account with balance & state. | Layer 2 off-chain solutions. | Public key transaction/ Merkle trees | Implement strong Nakamoto-Style consensus | Very expressive. E.g. Solidity. |
| eUTXO | Local and deterministic state. Maintain contract state. | Private-Key Wallet address | Outputs can be easily processed in parallel. | Public key transactions | Implement strong Nakamoto-Style consensus | More expressive than UTXO model. |

Fig. 9: State Transaction Model Comparison

## V. FINDINGS AND EVALUATION

Through this research we've observed the inability to successfully implement layer-one sharding on account based transactional models. As demonstrated in section III. B account based systems implement layer-two scaling to emulate the parallelization of UTXO sharding. We also observe the inability for off-chain automation dependent solutions to exist on UTXO based systems, due to smart contract incompatibility. The impact of this lack of solution interoperability between layer one and two presents a glass ceiling of scale out potential.

Layer two projects such as Plasma still rely on a heavily constrained main chain for coordination and dispute resolution, zero-knowledge and optimistic rollups all have similar dependencies on parent coordination. If sharding were implemented harmoniously underneath these architectures, bottlenecks would be avoided in main-chain coordination. This allows both scalability layers to scale-out by a factor of the other.

Imagine a scenario where a mainchain can handle 10,000 transactions-per-second through off-chain scaling and smart contract automation. The overall network architecture can continually scale on layer-two through nested parent/child relationships, however coordination and resolution cause a mainchain bottleneck. Resolving this with layer-one solutions, such as OmniLedgers linear scale-out sharding would increase potential speeds by a factor of n, where n is equal to mainchain parrelization.

To address this we propose the use of an Extended-UTXO model that allows for smart contract compatibility. This specification allows for off-chain automation in rollups, while maintaining the ability to shard UTXO states and transactions at layer-one.

## A. Design Architecture

The eUTXO follows the same graph based approach that the UTXO model does, complementing the concurrent and distributed attributes of blockchains. The graph based model eliminates the need for a globally shared mutable state, which is known to cause highly complex problems when facing computations involving that shared state [6]. The first account model blockchain, Ethereum, did solve how to run expressive smart contracts on a global state machine capable of execution on-chain. The eUTXO is an extension onto the UTXO ledger which notably increases the expressiveness of the model, while still utilizing the concurrent data flows of unspent transactions.

In creating a more extensible model, transactions can occur on more expressive state machines without needing to use an account model. Each individual transaction that occurs on a concurrent chain represents steps in state machine and to utilize a more expressive model, the state needs to be maintained and the same contract code is used along the whole sequence of transactions.

The eUTXO model maintains contract state by extending unspent transaction outputs from being a pair of validator and digital asset to a now being a treble containing a validator, the asset and a datum, where the datum contains arbitrary contract specific data [6]. The eUTXO model implements local state machines which are capable of running smart contracts on them to produces new unspent transactions as well as maintain the new state to be maintained. This enables developers to build distributed applications on massively scalable platforms.

OmniLedger inspired sharding protocols will be adapted to the eUTXO model using verifiable random function (VRF) to accomplish network sharding at layer-one [20]. We intend to use a temporary shard coordinator delegated through VRF election, to generate an unbiased random string. Each epoch the unbiased random string is used in shard assignment.

eUTXO transactions will be assigned through the use of a randomly chosen directory service shard. This phase is based on Zilliqa's approach to transaction sharding [22]. We intend to use the unbiased random string generated by the temporary epoch coordinator to delegate this directory service shard.

The directory shard will assign transactions to shards based on the sender's address and smart contract state being interacted with. Smart contract states will be maintained in the eUTXO transactions respective shard. Contracts will be fully executed globally at the end of each epoch when the directory service coordinator synchronises the entire network state.

As the eUTXO model supports the use of smart contracts, further layer 2 solutions such as Zero-Knowledge and Optimistic Rollups could be implemented, and even create nested-chains that communicate through smart contract two-way pegs which could increase throughput exponentially.

## VI. CONCLUSION

This research paper explored multiple scalability solutions both at layer-one and layer-two of the stack. The topics explored looked at how different protocols intend to increase the transaction throughput for a blockchain network and a comparative examination of these solutions was performed. The analysis was contextualised their feasibility in different use cases, and system implementations. Each scalability aspect explored was analysed regarding how they balanced security and decentralisation from the blockchain trilemma, while improving the horizontal scaling of systems.

We observed that there was a lack of scalability integrations across different layers that was attributed to horizontally scaling on both the layer-one with state sharding and and layer-two with smart contract implementations such as rollups or nested-chains. Most research identified was surrounding account model scalability, and very minimal work was done to explore smart contracts can be ran in parallel with UTXO.

We outlined a solution using making use of an extended and more expressive UTXO model, supporting paralellized UTXO state sharding, and support for smart contract dependant scaling through Zero-Knowledge rollups and nested side-chains.

### A. Future Works

We intend to implement a layer one test-chain using a eUTXO state transaction model, adapting OmniLedger inspired protocols to it. Smart contract integration and testing will occur once this is complete, and the viability of various approaches and architectures will assessed.

## REFERENCES

[1] J. Dattani and H. Sheth, "Overview of blockchain technology." *Asian Journal of Convergence in Technology*, pp. 123 – 140., 2019.
[2] Q. Zhou, H. Huang, Z. Zheng, and J. Bian, "Solutions to Scalability of Blockchain: A Survey," *IEEE Access*, vol. 8, pp. 16 440–16 455, Jan 2020.
[3] W. T. Jiacun Wang, *Formal Methods in Computer Science*. Andover, England, UK: Taylor & Francis, Jul 2019.
[4] L. N. Nguyen, T. D. T. Nguyen, T. N. Dinh, and M. T. Thai, "OptChain: Optimal Transactions Placement for Scalable Blockchain Sharding," in *2019 IEEE 39th International Conference on Distributed Computing Systems (ICDCS)*. IEEE, Jul 2019, pp. 525–535.
[5] L. Brunjes and M. Gabbay, "UTxO- vs account-based smart contract blockchain programming paradigms," Mar 2020. [Online]. Available: https://www.semanticscholar.org/paper/UTxO-vs-account-based-smart-contract-blockchain-Br%C3%BCnjes-Gabbay/ea415af5a99f27866d4a74601187d10a538fea50
[6] M. M. Chakravarty and J. Chapman, "The Extended UTXO Model," Feb 2020, [Online; accessed 20. Jan. 2022]. [Online]. Available: https://iohk.io/en/research/library/papers/the-extended-utxo-model
[7] M. Bellare, S. Halevi, A. Sahai, and S. Vadhan, "Many-to-one trapdoor functions and their relation to public-key cryptosystems," in *Advances in Cryptology - CRYPTO '98*. Berlin, Germany: Springer, May 2006, pp. 283–298.
[8] A. M. Antonopoulos and G. Wood, *Mastering Ethereum*. Sebastopol, CA, USA: O'Reilly Media, Inc., Nov 2018. [Online]. Available: https://www.oreilly.com/library/view/mastering-ethereum/9781491971932/ch04.html
[9] V. Buterin, "Ethereum whitepaper," *ethereum.org*, 2021. [Online]. Available: https://ethereum.org/en/whitepaper
[10] F. Santini, S. Bistarelli, I. Mercanti, and P. Santancini, "End-to-End Voting with Non-Permissioned and Permissioned Ledgers," *Journal of Grid Computing*, vol. 17, no. 3, Mar 2019.
[11] S. Nakamoto, "Bitcoin: A peer-to-peer electronic cash system," 2008. [Online]. Available: https://www.ussc.gov/sites/default/files/pdf/training/annual-national-training-seminar/2018/Emerging_Tech_Bitcoin_Crypto.pdf
[12] J. Lee, "Dappguard : Active monitoring and defense for solidity smart contracts," 2017. [Online]. Available: http://courses.csail.mit.edu/6.857/2017/project/23.pdf
[13] C. F. Tavares, B. and A. Restivo, "A survey on blockchain technologies and research." *J. Inf,*, pp. 118–128, 2019.
[14] R. D. L. S. Buterin, V. and G. Piliouras, "Incentives in Ethereum's hybrid Casper protocol." *International Journal of Network Management, 30(5)*, p. e2098, 2020.
[15] W. L. O. S. A. M. Tseng, L. and J. Othman, "Blockchain for managing heterogeneous internet of things: A perspective architecture." *IEEE network, 34(1),*, pp. 16–23., 2020.
[16] S. M. N. C. M. K. Dhumwad, S. and S. Prabhu, "A peer to peer money transfer using SHA256 and Merkle tree." *In 2017 23RD Annual International Conference in Advanced Computing and Communications (ADCOM)*, pp. 40 – 43., 2017.
[17] J. Frankenfield, "What is a merkle root ?" Aug 2021, [Online; accessed 29. Nov. 2021]. [Online]. Available: https://www.investopedia.com/terms/m/merkle-root-cryptocurrency.asp
[18] N. V. Z. C. B. K. G. S. Luu, L. and P. Saxena, "A secure sharding protocol for open blockchains," *2016 ACM SIGSAC Conference on Computer and Communications Security*, pp. 17–30, 2016.
[19] V. J. P. D. Goldberg, S. and L. Reyzin, "Verifiable random functions (VRFs).." ,p. ., 2018.
[20] E. Kokoris-Kogias, P. Jovanovic, L. Gasser, N. Gailly, and E. Syta, "OmniLedger: A Secure, Scale-Out, Decentralised Ledger via Sharding," in *2018 IEEE Symposium on Security and Privacy (SP)*. IEEE, May 2018, pp. 583–598.
[21] Z. S. X. G. G. Y. L. Y. X. J. Xi, J. and X. Zhang, "A Comprehensive Survey on Sharding in Blockchains. Mobile Information Systems," *Mobile Information Systems, 2021.*, 2021.
[22] Z. Dong, L., "The Zilliqa Project: A Secure, Scalable Blockchain Platform." ., p. ., 2018.
[23] K. Y. Kim, S. and S. Cho, "A survey of scalability solutions on blockchain." *In 2018 International Conference on Information and Communication Technology Convergence (ICTC)*, pp. 1204–1207., 2018.
[24] C. K. P. R. Z. Q. D. A. Singh, A. and K. Choo, "Sidechain technologies in blockchain networks: An examination and state-of-the-art review." *Journal of Network and Computer Applications,*, p. 102471., 2020.
[25] J. Poon and V. Buterin, "Plasma: Scalable autonomous smart contracts," *plasma.io*, 2017. [Online]. Available: http://plasma.io/plasma.pdf
[26] C. Sguanci, R. Spatafora, and A. M. Vergani, "Layer 2 Blockchain Scaling: a Survey," *arXiv*, Jul 2021. [Online]. Available: https://arxiv.org/abs/2107.10881v1
[27] "An Incomplete Guide to Rollups," Jan 2021, [Online; accessed 20. Jan. 2022]. [Online]. Available: https://vitalik.ca/general/2021/01/05/rollup.html

IEEE IOTJ., VOL. XX, NO. X, JANUARY 2022 9

[28] T. Schaffner, "Scaling public blockchains — a comprehensive analysis of optimistic and zero-knowledge rollups," 2021. [Online]. Available: https://wwz.unibas.ch/fileadmin/user_upload/wwz/00_Professuren/Schaer_DLTFintech/Lehre/Tobias_Schaffner_Masterthesis.pdf

[29] F. Bruschi, M. Tumiati, V. Rana, M. Bianchi, and D. Sciuto, "A scalable decentralized system for fair token distribution and seamless users onboarding," *Blockchain: Research and Applications*, p. 100031, Oct 2021.

[30] "Optimism Docs | Optimism Docs," Jan 2022, [Online; accessed 20. Jan. 2022]. [Online]. Available: https://community.optimism.io

[31] "Arbitrum Developer Quickstart · Offchain Labs Dev Center," Jan 2022, [Online; accessed 20. Jan. 2022]. [Online]. Available: https://developer.offchainlabs.com/docs/developer_quickstart

[32] "Introduction to zkSync for Developers | zkSync: secure, scalable crypto payments," Jan 2022, [Online; accessed 20. Jan. 2022]. [Online]. Available: https://zksync.io/dev

[33] "StarkEx Introduction," Jan 2022, [Online; accessed 20. Jan. 2022]. [Online]. Available: https://docs.starkware.co/starkex-v4

[34] S. Johnson, P. Robinson, and J. Brainard, "Sidechains and interoperability," *arXiv*, Mar 2019. [Online]. Available: https://arxiv.org/abs/1903.04077v2

[35] D. T. L. D. C. E. L. Q. Dang, H. and B. Ooi, "Towards scaling blockchain systems via sharding." *In Proceedings of the 2019 international conference on management of data*, pp. 123 – 140., 2019.

[36] H. A. Hafid, A. and M. Samih, "New mathematical model to analyze security of sharding-based blockchain protocols," *IEEE Access.*, pp. 185 447–185 457., 2019.

[37] F. Sanchez, "Cardano's Extended UTXO accounting model – built to support multi-assets and smart contracts (part 2) - IOHK Blog," Mar 2021, [Online; accessed 21. Jan. 2022]. [Online]. Available: https://iohk.io/en/blog/posts/2021/03/12/cardanos-extended-utxo-accounting-model-part-2